%% file: main.tex
\documentclass{article}
\usepackage{fixltx2e,fix-cm}
\usepackage{amssymb}
\usepackage{amsmath}
\usepackage{graphicx}

\usepackage{makeidx}
\usepackage{multicol}
\usepackage{setspace}
\usepackage{subcaption}
\usepackage[export]{adjustbox}

\makeindex

\usepackage{fullpage}
\usepackage{bm}
\usepackage{graphicx,psfrag,epsf}
\usepackage{xcolor}

\title{An Introduction to the Calibration of Computer Models}

\author{Richard D. Wilkinson, Christopher W. Lanyon}

\begin{document}
\include{preamble} 

\maketitle

\section{Introduction}
In the context of computer models, calibration is the process of estimating unknown simulator parameters from observational data. Calibration is variously referred to as model fitting, parameter estimation/inference, an inverse problem, and model tuning.
The need for  calibration
occurs in most areas of science and engineering, and has been used to estimate hard to measure parameters in models of climate \cite{bellprat2012objective,chang2019computer,couvreux2021process, guzman2010genetic,hourdin2021process, hourdin2017art, chang2016calibrating, williamson2013history}, cardiology \cite{caruel2014dimensional,marchesseau2013fast,plumlee2016calibrating,rodero2023calibration,whittaker2020calibration}, drug therapy response \cite{frieboes2009prediction,huang2019tuning}, hydrology \cite{arsenault2014comparison,bardossy2007calibration,gupta1998toward, beven2001equifinality} and many other disciplines. Although the method of calibration can vary substantially, the underlying approach is essentially the same and can be considered abstractly.
To set notation,  denote the unknown parameter that is to be estimated as $\bx \in \mathcal{X}$, and consider the computer model $f$ to be a map from $\X\times \U$ to output space $\F$
$$ f: (\bx, \bu) \mapsto f(\bx, \bu)\in \F.$$
$\X, \U$, and $\F$ may  be multidimensional, and will often (but not necessarily) be subsets of Euclidean space.
Here, $\bu \in \U$ denotes a control input (which may not be present in some simulators), which is a setting in the simulator that specifies
inputs that do not need to be estimated, such as experimental conditions
under which the data were collected.
We will sometimes drop $\bu$ from the notation and just write $f(\bx)$. The observational data we will use  to calibrate the simulator is $\by \in \Y$, where $\Y$ is an n-dimensional space. If $\bu^*$ denotes the control inputs/experimental conditions under which $\by$ was collected\footnote{We may have several datasets $\by^{(1)}, \ldots,\by^{(K)}$ collected under conditions $\bu^{(1)}, \ldots, \bu^{(K)}$. The aim is to find $\bx^*$ such that $f(\bx^*, \bu^{(k)})$ explains $\by^{(k)}$.}, then we assume that the simulator can in some sense {\it explain} $\by$ if we find the correct input parameter value $\bx^*$. In other words,  for some $\bx^*\in\X$, $f(\bx^*, \bu^*)$ relates to $\by$ in a sense that will be made precise in the next section. Calibration is the process of estimating $\bx^*$ from $\by$.

\begin{figure}[h]

     \begin{subfigure}[t]{0.3\textwidth}
         \includegraphics[width=\textwidth,valign=t]{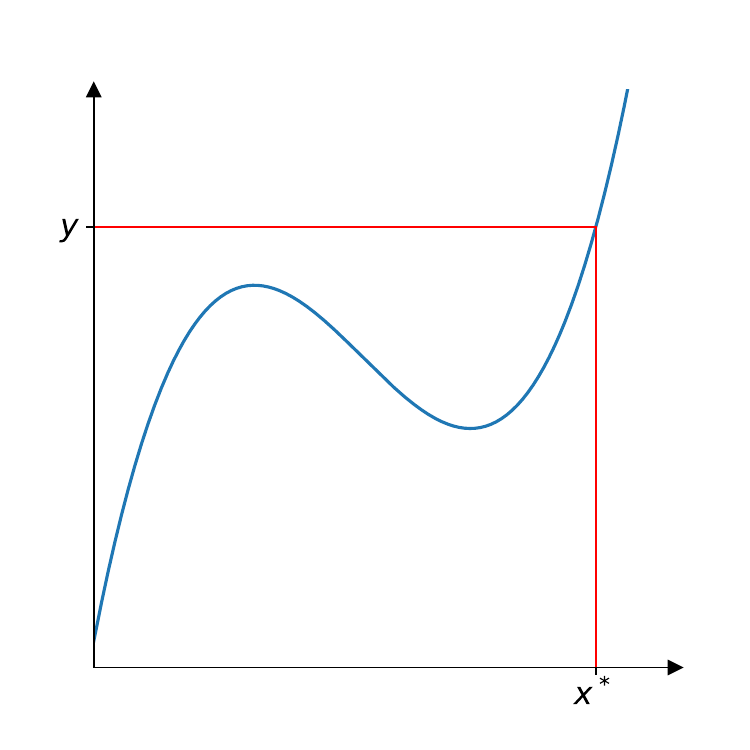}
         \caption{}
         \label{fig:yObs}
     \end{subfigure}
     \hfill
     \begin{subfigure}[t]{0.3\textwidth}

         \includegraphics[width=\textwidth,valign=t]{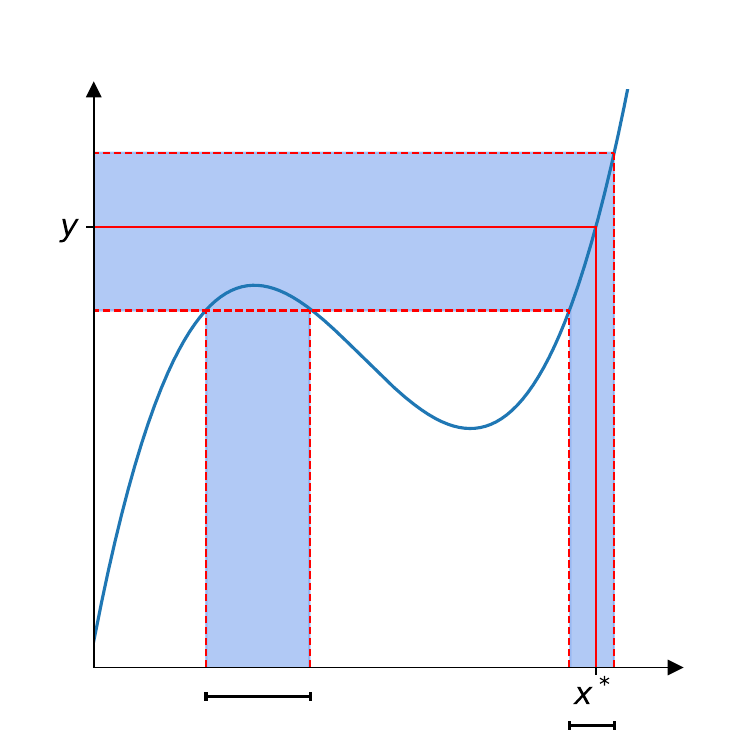}
         \caption{}
         \label{fig:yObsNoisy}
     \end{subfigure}
     \hfill
    \begin{subfigure}[t]{0.3\textwidth}
    \includegraphics[width=\textwidth,valign=t]{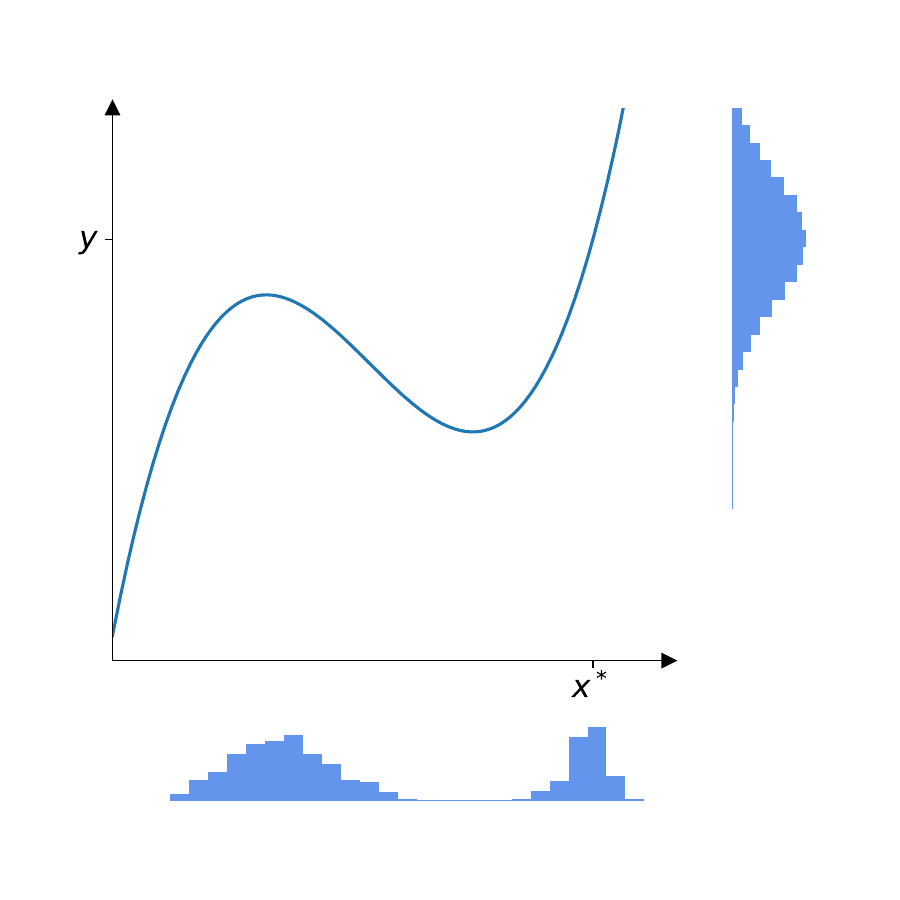}
         \caption{}
         \label{fig:yObsGauss}
     \end{subfigure}
        \caption{The blue line shows the functional relationship $y=f(x)$ representing the simulator. Plot (a) shows the case where we have a perfect observation (i.e., with no uncertainty), and the red lines show how $y$ maps down to the best input $x^*$. Plot (b) shows how an interval on the y-axis maps to two intervals of feasible parameters on the $x$-axis. Plot (c) shows probabilistic calibration, where we assume  Gaussian noise on the observation, and show how this translates to a statistical distribution for $x^*$.}
        \label{fig:calib1}
\end{figure}

Figure \ref{fig:calib1} illustrates this process in one dimension.  Each panel  shows the  functional relationship  $y=f(x)$ as a solid line,  representing the simulator output.  Panel \ref{fig:yObs} shows the simplest  situation where we have a precisely observed value of $y$. We can then read across and down to see what value of $x$, denoted $x^*$, leads the simulator to predict $y$. In panel \ref{fig:yObsNoisy}, we assume  we observe $y$ with error: the observation is shown as an interval on the $y$-axis, which leads to two disconnected regions of input space for values of $x$ consistent with the observations. Panel \ref{fig:yObsGauss} shows the case where we assume Gaussian observation error, that is, data is generated as $y=y^{true}+N(0, \sigma^2)$, which if we then read across and down,  translates to a distribution on the $x$-axis. 

In this chapter, we give  a brief introduction and guided tour of calibration methodology,  organised by the key decisions that must be made when calibrating a simulator:
\begin{itemize}

\item \textit{Observational model}: how does the simulator output relate to the data? We need a statistical model for the measurement error and variability inherent in the data, but we must also consider the accuracy (or otherwise) of the simulator. Is the simulator a perfect representation of reality, or is it misspecified to some degree? If the latter, how do we represent the discrepancy between model and reality?

\item  \textit{Calibration framework}: what quantity are we trying to find? Is it  a point estimate, such as  a maximum likelihood estimator, a frequentist confidence interval, or a Bayesian (or Bayes-like) posterior distribution? Our focus is on the latter, but we will touch on methods for the former.

\item  \textit{Quantity of interest}: why are we calibrating the model? Is it because we are interested in the true physical value of the input $\bx^*$ given observation $\by$, or is it that we want to do calibrated prediction of some future $\by$ given the historic data? Or do we want to use the simulator to inform a decision, for example, about the best way to further develop the model? Our aim may  inform our approach to calibration.

\item {\it Computational approach}:  will we  take an optimization or a sampling based approach to calibration? And what degree of accuracy do we require in our approximation to the solution? In most problems, there will be constraints on the computational budget available for calibration, which will influence our choice.

\end{itemize}

\noindent These decisions are a matter of application specific judgement, but also depend on the properties of the simulator.
In this chapter, we will discuss these points, and highlight methods and approaches that can be used for  calibration. It is not an exhaustive review, but  a guided tour illustrating a range of calibration methodology.







\section{Observation model}

Perhaps the most important and difficult decision is deciding upon an observation model, namely, stating how we believe the simulation of some phenomenon relates to the observed data. Observation models are usually statistical in nature, but if  detailed mechanistic information about the observation process is available, then this should be  incorporated \cite{carson2019quantifying}. The \textit{best-input} approach assumes that for some $\bx^*\in\mathcal{X}$, the simulation $\bbf(\bx^*, \bu)$ gives the best prediction the simulator is capable of for control input $\bu$. The observation model should  describe how this best possible simulator prediction relates to the observations $\by$.
Note that the best input approach assumes there is a single $\bx^*$ for all choices of the control inputs $\bu$, usually because we expect $\bx^*$ to correspond to  some true but unknown physical constant. The observation model  defines the statistical  likelihood function, $\pi(\by \mid \bx)$, which can be thought of as the probability of seeing data $\by$ when using input $\bx$, and it is often used to calibrate the models. When $\bbf$ is a deterministic simulator, $\pi(\by \mid \bx)\equiv\pi(\by \mid \bbf(\bx))$, is straight-forwardly defined by the observation model. But when $\bbf$ is a stochastic simulator,  there is an additional step to compute the likelihood:
\begin{equation}
\pi(\by \mid \bx)=\int \pi(\by \mid \bbf(\bx, \bu))\pi (\bbf(\bx,\bu) \mid \bx) {\rm d} \bbf.
\label{eqn:likelihood}
\end{equation}
The first term in the integral is the observation model we must define; the second term is the simulator distribution when run at $\bx$, which may be unknown making the likelihood intractable.



Simulators often output more information than is present in the data. For example, $f$ may simulate an entire spatial-temporal trajectory, but we may only have observations at particular points in space and time. Or perhaps we only observe some of the modelled variables (e.g., observe location but not velocity). Or the observation process may integrate the output over some time period (e.g., historic climate observations are often of the form of cumulative rainfall \cite{rivero2013time} or growing degree days \cite{mcmaster1997growing}: measurements that integrate precipitation or temperature). Another common situation is where we look for emergent behaviour in the output, for example, climate simulators output daily weather patterns and we may then identify the occurrence or otherwise of weather phenomena such as El Ni\~{n}o.  More generally,  we may examine the simulator for  physically realistic behaviours (sometimes called {\it precalibration}, see \cite{edwards2011precalibrating}). We can represent all of these situations by an observation operator, $\bg: \F \rightarrow \Y$, that acts on the simulator output to produce a single directly comparable output for each of the $n$ observations in the vector $\by$:


\begin{equation}
\bg: \bbf(\bx, \bu) \mapsto \bg(\bbf(\bx, \bu))=
\left( g_1(\bbf(\bx, \bu)),
\;\ldots, \;
g_n(\bbf(\bx,\bu))\right)^\top.
\end{equation}
Note that this allows for each output to be a different type of data (continuous, discrete, categorical etc).
The choice of $g_i$ will depend on the format of our data and the output of our model, $\bbf$. For example, in \cite{gahungu2022adjoint}, it is assumed that observations are generated by space and time averaging over a small part of the domain of an advection-diffusion regime. In that case $\bbf$ is the solution of the advection-diffusion equations and $g_i$ encodes the space and time averaging. Of course $g_i$ could encode any relation between the model output and observations, but at its simplest, $g_i$ could simply be the identity function.

We must then choose an observation model, which  states how $g_i(\bbf(\bx^*, \bu^*))$ relates to the observation $y_i$.
For some observations, we may choose to assume that the best simulation should perfectly match  the data, i.e., that $y_i = g_i(\bbf(\bx^*, \bu^*))$.
This is unlikely to be true for any continuous quantity, but may be a reasonable assumption for discrete or categorical outputs, such as  the presence or absence of some property.
For other observations, we may assume that the best simulator prediction is unbiased, but that there is some noise in the data. In this case we may assume
\begin{equation}
y_i = g_i(\bbf(\bx^*, \bu^*))+\epsilon_i
\label{eq:obs}
\end{equation}
where, $\epsilon_i$ is an observation error term with zero mean. If we plan to do probabilistic inference of $\bx^*$, we need to specify distributions for the error terms; the most common choice is to assume a normal distribution,  $\epsilon_i \sim N(0, \sigma_i^2)$, but we can add as much complexity here as is justified. For example, adding correlations between different outputs (i.e., assume  $\operatorname{Cor}(\epsilon_i, \epsilon_j)=c_{ij}$), using heavier tailed or skewed distributions, 
or assuming other error structures such as multiplicative errors etc. If we can fully specify the observation error distribution, this will simplify the inference, but in many situations we will need to include free parameters in the statistical observation model (such as $\sigma_i^2$ above) and add these to the list of parameters to be inferred in the calibration.

The most complex situation, and also the most common, is where the simulator is misspecified in some way \cite{box1976science},
 that is, there is a bias so that even when run at the best input parameter $\bx^*$, the distribution of the simulated data (e.g. the right hand side of Eq. \ref{eq:obs}), does not match the distribution of the data generating process. One approach in this situation 
 is to labour under the assumption the model is correct (i.e., that Eq. \ref{eq:obs} holds), calibrate the simulator, and then look at what goes wrong and  attempt to learn from this to improve the simulation and observation model, before repeating the calibration. Note that it can be difficult to spot misspecification in many cases, but  looking at calibrated predictions (see Eq. \ref{eq:calibpred}) from the simulator will often illustrate the problem. Ignoring misspecification can  often  result in over-confident or physically implausible parameter estimates \cite{yang2018bayesian, ramamoorthi2015posterior, grunwald2017inconsistency}, and we may find that calibrating to different datasets, or parts of the data, leads to different and incompatible parameter estimates violating the best input assumption. For example, if we have data $\by^{(i)}$ collected under experimental conditions $\bu^{(i)}$, then we may find calibrating to $\by^{(1)}$ leads to a different and incompatible inference for $\bx^*$ than when using $\by^{(2)}$.

But what do we do if  we have gone through repeated cycles of model development, and are now faced with needing to use a still imperfect simulator to make some decision, prediction, or inference? There are statistical approaches that aim to be robust to misspecification, usually by abandoning the likelihood \cite{bissiri2016general, jewson2018principles, matsubara2022robust, grunwald2021pac, shafer2008tutorial} or adapting it in some way, for example, introducing some learning rate to the likelihood in order to aid convergence to the posterior distribution \cite{grunwald2017inconsistency}.
But these approaches often only work well in models that are slightly misspecified, for example, where there are heavier tails in the observation error, $\epsilon$, than assumed, or where there are outliers (so that the data generating process is accurately represented by the simulated process some large proportion of the time, but  occasionally includes observations drawn from a very different polluting distribution). In other words, when the simulator is essentially sound, but the observation model is at fault.
However, when working with complex simulators we often find there is a gross misspecification in some of the outputs. In this case, we can include a data-driven model of the way in which the simulator fails, $d$, called the discrepancy model and then learn the discrepancy at the same time as learning $\bx^*$ in the calibration process \cite{kennedy_bayesian_2001}. For example, a common choice is to assume
\[y_i = g_i(\bbf(\bx^*))+d_i+\epsilon_i\]
where $d$ is modelled with a flexible data driven model such as a Gaussian process \cite{rasmussen2006gaussian} or neural network  \cite{goodfellow2016deep}.
The details of how we model $d$ will depend on the situation, but as an example, suppose we are modelling a spatial process (such as temperature around the globe), so that the data and simulator are functions of space, i.e.
$\by\equiv y[\bw]$ and $\bbf(\bx) \equiv f(\bx)[\bw]$
where $\bw$ indexes space. The simulator discrepancy will then also be a function of space, $d[\bw]$, which we may choose to model with a  Gaussian process, i.e.,
$d[\bw]\sim GP(m(\bw), c(\bw, \bw'))$. If we collect observations at locations $\bw_1, \ldots, \bw_n$, then our observation model is
\begin{equation}y_i= y[\bw_i] =  f(\bx^*)[\bw_i] +d[\bw_i]+\epsilon_i. \label{eqn:modelwithdiscrepancy} \end{equation} The GP prior assumption for $d[\cdot]$ will give us the joint distribution of $(d[\bw_1], \ldots, d[\bw_n])^\top$, which can be used to compute the likelihood function.

Including a discrepancy term can  complicate the inference, in part because it often introduces a non-identifiability between $\bx^*$ and $d$  into the model \cite{arendt2012quantification}. To see why, note that in Eq. \ref{eqn:modelwithdiscrepancy} we  model $y[w]$ as the sum of two functions of $\bw$: $f(\bx)[\cdot]$ and $d[\cdot]$. If we use a flexible model for $d$, then there are many ways for $d$ to correct the predictions of $f$. Indeed, given any $\bx$, there is a $d[\cdot]$ that when added to $f(\bx)[\cdot]$ will correct it to match $y[\cdot]$.
When we jointly estimate $\bx$ and $d$, the assumptions that are (often implicitly \cite{gelman2017prior}) made in our prior distributions for $d$ and $\bx^*$, often determine what the posterior for $\bx^*$ will be \cite{brynjarsdottir2014learning}. Although  building in additional information about $d$ can help \cite{brynjarsdottir2014learning},  a useful rule-of-thumb  is to assume that although we can correct our predictions from $f$ by using a discrepancy model, it is much harder to correct  our inferences about $\bx^*$. This will be discussed further in the next section.



\section{Calibration framework}

How should we characterize what is a good or bad parameter value? How do we define the best parameter value? And how should we represent uncertainty about this value? These are all questions about what statistical framework to use.
The performance of parameter $\bx$ in the context of the simulator, is  measured by defining some score $S(\bx)\equiv S(\bx, \bbf, \by)$, which is a function of the parameter, the model prediction with that parameter, and the data\footnote{In this section, we  suppress the dependence on the control inputs, $\bu^*$ corresponding to $\by$ in the notation.}.
Parameters that lead to lower scores are judged to be better than parameters which result in higher scores.
For probabilistic models (i.e. where we specify distributions for error terms), the standard approach is to use a score based on the likelihood function, such as the negative log likelihood\footnote{We can use the log-likelihood (i.e. without the negative), but we then need to remember that bigger scores indicate better performance.} $S(\bx, \bbf, \by) = -\log \pi(\by \mid \bx)$.
The maximum likelihood estimator (MLE), $\hat{\bx}$, is found by minimizing this with respect to $\bx$ using standard optimization methods.
The MLE is known to have desirable statistical properties in well-specified models: consistency (it converges to the true value asymptotically as the number of data points grows), efficiency (it asymptotically achieves the minimal possible variance for an unbiased estimator), asymptotic normality (the distribution of $\hat{\bx}$ converges towards a Gaussian distribution as the amount of data grows) \cite{van2000asymptotic}, which can be used to compute approximate confidence intervals in simpler problems.

Although the log-likelihood (or some modification of  it, such as the log-likelihood plus the logarithm of a prior distribution) is the default choice of score for many statisticians, in some situations we may choose to use scores not based on the log-likelihood. For example, for stochastic  simulators the likelihood, Eq. \ref{eqn:likelihood} may be intractable, or it may be that we don't wish to specify a  statistical  distribution  for the error model (for example, in least squares regression we only make assumptions about the mean and variance of the errors).
One of the drawbacks of using the likelihood is that it can be  sensitivity to model misspecification, in which case it converges to the {\it pseudo-true} value, which is the value that minimizes the Kullback-Leibler divergence between the simulator and the true data generating process \cite{van2000asymptotic}.
This can be the case even for mild misspecification such as occasional outliers or heavier tails than assumed for the error model, and so  it can often be better to use simpler scores that compare the data to aspects of  the simulator that we are confident in \cite{bissiri2016general}.
For example, if there are multiple outputs to the simulator, then we often use a summary statistic \cite{prangle2015summary}, $\bT(\cdot)$, and then compare the summary of the output, $\bT(\bg(\bbf(\bx)))$ with the summary of the data,  $\bT(\by)$. If a simulator is misspecified, we may be able to find a summary $\bT(\bg(\bbf(\bx)))$ that is less misspecified than $\bg(\bbf(\bx))$, and thus which will give more reliable inferences. For example, a model with a periodic dynamic output may be out of phase with the data, but may accurately predict summaries of the output such as the period, amplitude or other emergent properties, and  thus may be a more sensible output to calibrate to than the raw time-series output.
Note that unless we are able to use sufficient statistics as summaries \cite{joyce2008approximately}, which are usually unavailable for complex problems, using a summary will incur a loss of statistical  efficiency in the well-specified case. But this is often a price worth paying when the model is misspecified as it can result in a lower bias and more reliable calibrations and predictions.


If  a point estimate of the parameters is all that is required, then once we have defined a score, we can simply choose  $\bx^*$ to minimizes the score
\[\bx^* =\arg\min_{\bx\in\mathcal{X}} S(\bx, \bbf, \by).\]
For flexible simulators, there can be a risk of `over-fitting' the data, which is where the simulator explains the noise as well as the signal, and it usually results in poor predictive performance on new data.
It can usually be identified by  splitting the data into test and training sets: fit the simulator using the training data, and evaluating the resulting calibrated simulator on the test data to check for over-fitting \cite{hastie2009elements}. When calibration is fast,  cross-validation can be used \cite{berrar2019cross}.

Bayesian approaches are the most commonly used framework for characterizing uncertainty about the estimated parameters, in part because of their conceptual simplicity, as we only need deal with probability distributions. They provide a coherent approach for combining difference sources of information, and  it can be relatively easy to compute approximations to the posterior distribution in many cases. The biggest drawback is that we are required to specify a prior distribution $\pi(\bx^*)$ describing the uncertainty about $\bx^*$ before seeing the data\footnote{This is also a strength of Bayesian methods, as it allows us to build expert knowledge into the analysis, strengthening the calibration.}, and for many problems we may find that our computed posterior distribution
$$\pi(\bx^* \mid \by)= \frac{\pi(\by \mid \bx^*) \pi(\bx^*)}{\pi(\by)}$$
is sensitive \cite{berger1990robust} to this choice (see, for example, chapter 5 of \cite{lunn2012bugs}).


Generalizations of the Bayesian approach that typically don't use the likelihood function to score the data, are becoming increasingly popular \cite{jewson2018principles, knoblauch2019generalized, bissiri2016general}.
Two closely related  variations on standard Bayes that are popular in the computer experiment literature, are history matching \cite{craig1997pressure, williamson2013history,vernon2014galaxy} and approximate Bayesian computation (ABC) \cite{csillery2010approximate,sisson2018handbook,sunnaaker2013approximate, marin2012approximate}.
In both \cite{holden2018abc}, the acceptable region of parameter space is defined by thresholding some score $S(\bx, \bbf, \by)$ at some tolerance $\tau$:  any $\bx$ with $S(\bx, \bbf, \by)< \tau$ is deemed acceptable. This can be a crude but effective way
to incorporate a simple discrepancy model, without needing to specify a full probability distribution for $d$ \cite{wilkinson2013approximate, holden2018abc}.

In history matching, an `implausibility' score is used \cite{craig1997pressure}, which is essentially the Mahalanobis distance between $\by$ and $\bbf(\bx)$ (or $\bT(\by)$ and $\bT(\bbf(\bx))$) as it  uses the squared difference between simulator and data  scaled this by  the variance  of the errors (including observation error, model discrepancy, and code uncertainty if a surrogate is used). Rules of thumb, based on the expected probability mass within three sigma of the mean \cite{pukelsheim1994three}
are commonly used to determine sensible values of the tolerance $\tau$. History matching is usually presented as a non-probabilistic approach, that simply classifies space as either implausible  or not, rather than giving a posterior distribution over $\mathcal{X}$

In ABC, arbitrary scores are used, and the threshold $\tau$ is often set by computational constraints, for example, by accepting the best 1\% of simulations from some large ensemble. ABC is most commonly used in the setting of stochastic simulators, and so some simulations with a given $\bx$ will be accepted and some rejected, with higher acceptance rates for the best input $\bx^*$. This, and the incorporation of prior information about $\bx$, allows  ABC  to give a probabilistic posterior distribution that approximates the standard Bayesian posterior, becoming more accurate as the tolerance $\tau$ gets closer to $0$.  Although ABC was initially proposed \cite{beaumont2002approximate,tavare1997inferring} for situations in which the simulator likelihood function is unknown but can be sampled from, it has found applications in situations where the likelihood is known (and so standard Bayes could be used).
In part, this is because
ABC can be interpreted  as  giving Monte Carlo exact inference, but for a model that assumes some  level of simulator misspecification \cite{wilkinson2013approximate}.

There are also frequentist approaches to  calibration which seek to find confidence intervals for $\bx$, rather than a distribution over the parameter values \cite{joseph2009statistical}.
Often, the model discrepancy term, $d$, is ignored and maximum likelihood estimates of $\bx^*$ are generated via evaluations of $\bbf(\bx)$ (or surrogate of $\bbf$) during some optimisation procedure or parameter search (see, e.g., \cite{vecchia1987simultaneous, huang2006global, vanni2011calibrating}).  In \cite{wong2017frequentist}, a method is presented that incorporates a non-parametric representation of the model discrepancy term and estimates $\bx^*$ by solving a minimisation problem, with $d$  subsequently estimated using non-parametric regression between the observations and model outputs, with confidence intervals estimated via bootstrapping. Identifiability for $\bx^*$ and $d$  can be recovered by redefining $\bx^*$ as a pseudo-true value  that minimizes the simulator mean square error. As in similar Bayesian approaches \cite{plumlee2017bayesian, tuo2015efficient}, this  changes the interpretation of the parameters $\bx$.


\section{Quantity of interest: why are we calibrating?}

Before deciding how to calibrate a simulator, and investing time in developing a detailed observation model, it helps to be clear about why we are calibrating. Common motivations include

\begin{enumerate}
    \item Calibrated prediction
    \item Parameter inference
    \item Scientific understanding/model development.
\end{enumerate}

In a Bayesian setting, calibrated prediction aims  to find the prediction of the model after marginalizing (i.e. integrating) out uncertainty about the parameters.
For example, suppose we collect data $\by$ under experimental conditions $\bu^*$, and then want to use this to predict what data, $\by^p$ say, we will see under conditions $\bu^p$. In a Bayesian framework, the calibrated prediction is given by
\begin{equation}
\pi(\by^p \mid \by, \bu^*, \bu^p) = \int \pi(\by^{p} \mid \bx^*, \bu^p) \pi(\bx^* \mid \by, \bu^{*}) {\rm d} \bx^*.
\label{eq:calibpred}\end{equation}
If the computational budget allows us to evaluate the simulator sufficiently often, then we can estimate this by forward propagating a representative sample of parameters, $\bx^{(1)}, \ldots, \bx^{(N)}$ sampled from the posterior.
In misspecified models, we need the discrepancy model $d$ to correct the observation model in such a way  that it allows us to predict well \cite{brynjarsdottir2014learning}, which can be   tested using held-out data.

The situation where  interest primarily lies in  parameter inference, i.e., in $\pi(\bx \mid \by)$, can be somewhat harder than calibrated prediction when we have a  misspecified simulator. This may seem counter intuitive at first, as we require the posterior to compute the Bayesian posterior predictive distribution given in Eq. \ref{eq:calibpred}. The reason it can be more challenging is that we don't just need to correct the model so that it predicts well, but correct it so that $f(\bx)$ mechanistically reproduces  real world behaviour. When we care about some physically defined real quantity, `$x$', we need the model to accurately represent the link between this real quantity `$x$' and the data. We may have a quantity in the simulator that we refer to using the same label `$x$' as the physical quantity, but when a simulator is misspecified, it can subtly change the interpretation so that the simulator's `$x$' no longer corresponds to the physical quantity `$x$'.
A striking example of this occurs in global circulation models of the climate, where  a much larger value of `viscosity' is required in the simulator than is  physically realistic \cite{large2001equatorial}. In other words, what is referred to as `viscosity' in the simulator is a different quantity to the viscosity of the ocean. The effect is due to the discretization of the underlying equations and is
well understood in  this case, but if we naively used the simulator to infer viscosity we would end up with a meaningless inference in terms of the true viscosity.

Fixing this  requires us to fix the simulator so that the functional link between $x$ and $y$ is correct. In this case, and when  scientific understanding or model development is our goal, there are a variety of approaches we can take. One option is to fit the simulator with no discrepancy model, i.e. use Eq. \ref{eq:obs}, and then  seek to understand what has gone wrong by looking at where the simulator is unable to fit the data well, where it predicts badly, known biases in estimates of $\bx^*$ etc. Or we can fit a discrepancy model, $d$, and then look at the inferred discrepancy   for clues as to how to improve the simulator. Note that when working with only the output of the simulator, it can be difficult to understand the fitted discrepancy model $d$. For example, if our simulator is based on the differential equations, $\rd y/\rd t=h(x,u, y,t)$, then learning the error in $h$ from the solution space $y(t)$ can be difficult. If we can instead access the  internal simulator state, then (at least for simpler models) we can try to fix the underlying equations $h$ by including a discrepancy term on the derivatives (e.g. $\rd y/\rd t=h(x,u,y,t)+d(u, y,t)$ \cite{wilkinson2011quantifying,raissi2018hidden, schaeffer2017learning, chen2018neural, raissi2019physics}, but this can be computationally challenging.




\section{Computational approach}

Once we have decided upon an observation model and a calibration framework, we can then focus on how to do the computation necessary to find the quantities of interest. For complex simulators, we will rarely be able to  derive these quantities mathematically, and instead will  need to choose a numerical scheme to approximate them.
The choice of computational approach will  depend on several key considerations. Firstly, how computationally expensive is it to evaluate the simulator and what is the available computational budget?
Furthermore, is this resource available in parallel (so that multiple simulations can be run simultaneously) or sequentially?
If your computer model takes a fraction of a second to evaluate, there will be many viable approaches to calibration. But  for expensive simulators,   we will need to choose a method that uses just a small ensemble of simulator evaluations, such as surrogate model methods.
The choice will also depend on how much time we are prepared to spend on the approximation.  If we need to calibrate very quickly (e.g., in close to real time), we will need a very different approach (such as an amortized inference network \cite{gershman2014amortized}) to when we are willing to spend weeks computing the most accurate inference we can.

Our knowledge about the simulator should also inform our analysis.
For example, is the output a continuous smoothly varying  function of the input parameters $\bx$? If so, na\"{\i}ve sampling methods often ignore this and have to relearn it during the inference \cite{rasmussen2003gaussian, wilkinson2014accelerating}.
Can we exploit any known structure in the simulator, such as symmetries \cite{uteva2017interpolation, bloem2020probabilistic, ginsbourger2016degeneracy}?
Is the simulator stochastic or deterministic? If we know the simulator essentially solves a set of mathematical equations (such as a set of differential equations),
and if we are able (and willing) to modify the computer code, there are often intrusive calibration approaches that may be efficient in some settings \cite{goh2013prediction, march2012constrained}. Otherwise we are in a black-box situation where for any given input $\bx$, we are only able to query the simulator to find $f(\bx)$. We can classify black-box calibration methods into {\it zeroth-order} methods, that use only the simulator response $f(\bx)$, {\it first-order} methods which additionally use gradient information $\nabla f(\bx)$, and {\it second-order} methods which use the Hessian/curvature information $\nabla^2 f(\bx)$ etc.  Generally higher order methods will result in more accurate approaches for a given cost. Although derivative information about simulators is increasingly available to modellers, either via adjoint methods \cite{estep2004short,cao2003adjoint}, or via automatic differentiation software \cite{abadi2016tensorflow, paszke2017automatic, jax2018github, carpenter2017stan}, it is still often unfortunately the case that for  complex simulators  gradient information willbe  unavailable.


Finally, we must also take the dimension of $\bx$ into consideration, as well as how much data is available to us. For low dimensional problems, we may be able to use numerical quadrature methods \cite{davis2007methods} to compute very accurate approximations to the posterior distribution, whereas high dimensional problems may be intractable unless we can find correlation/structure in $\bx$ that allows the dimension of the problem to be reduced in some way \cite{constantine2015active, constantine2016accelerating, zahm2022certified,cui2014likelihood}. If we have large amounts of data, then there may be computational approaches that allow us to efficiently approximate the posterior \cite{bierkens2019zig}, conversely, in problems with only sparse data, we may need to focus  on more carefully eliciting a prior distribution for $\bx^*$ \cite{o2006uncertain,gosling2018shelf}.

\subsection{Sampling approaches}

The default approach for Bayesian problems, at least when using computationally cheap simulators, is often to use some form of Monte Carlo sampling to approximate the posterior distribution \cite{robert1999monte}. For example, Markov chain Monte Carlo (MCMC) methods \cite{brooks2011handbook}  generate a random sequence of parameter values $\bx^{(1)}, \bx^{(2)}, \ldots$, which eventually
will converge to being a sample from the posterior distribution $\pi(\bx^* \mid \by)$. The Metropolis-Hastings algorithm is the simplest approach,  and in theory, all it requires is for the user to define a proposal distribution $q(\bx' \mid \bx^{(n)})$ describing how to generate moves from the current state $\bx^{(n)}$, to a newly proposed state $\bx'$. The algorithm then decides whether to accept or reject that move depending on the basis of the relative likelihood and prior support of $\bx'$ and $\bx^{(n)}$. The difficulty comes in finding a proposal distribution that generates a Markov chain $\bx^{(1)}, \bx^{(2)}, \ldots$, that converges rapidly, and which fully explores the posterior support (i.e., that \textit{mixes} well in the jargon of MCMC). Once such a proposal is found, we can then generate sufficient samples to approximate the posterior distribution (for example, by using a kernel density estimator) or any posterior expectation \cite{dellaportas2003introduction}. In practice, it can be hard to find a good proposal distribution, particularly  when  the posterior distribution is multi-modal.

Many specialised MCMC algorithms have been developed that use different forms for the proposal distribution, and these can greatly improve sampler performance, but typically require additional knowledge or structure which may not be available for complex problems.
Zeroth order methods include Gibbs sampling \cite{gelfand1990sampling}, which requires the full conditionals, $\pi(\bx_{[i]} \mid \bx_{[-i]}, \by)$ (which  may be available for parameters in the observation model if conjugate priors are chosen, but not usually for simulator parameters); slice sampling \cite{neal2003slice}; as well as algorithms specifically designed for expensive simulators with certain structures  \cite{dodwell2015hierarchical, christen2005markov, cotter2013mcmc}.
If gradient information is available, then this can be exploited to speed up convergence and mixing so that fewer simulations are required \cite{neal2012bayesian, hoffman2014no, roberts1998optimal, girolami2011riemann}, with Hamiltonian Monte Carlo (HMC) algorithms (such as those used in the probabilistic programming languages   STAN \cite{carpenter2017stan}) being particularly successful at exploring posterior distributions with  highly correlated parameters.
Algorithms can often be combined so that structure is exploited for the parameters for which it is available.

The advantages of MCMC are that it is relatively easy to code, can require little knowledge of the simulator (e.g.  Metropolis-Hastings and slice sampling), and in theory at least, will converge to arbitrary accuracy  asymptotically, although the value of this is somewhat limited in practice.
Drawbacks  include that it is sequential in nature making distributed computation (i.e., on parallel architectures) challenging,  it can be hard to assess convergence and mixing (i.e.,  whether a chain it has explored the entire parameter space, particularly in multi-modal posteriors), but primarily, that the number of simulator evaluations required will be infeasible for many problems\footnote{Even for simple problems, it is common to require tens of thousands of simulations. It is not just the simulations needed for the final production MCMC run, but all of the simulations required to develop a  good proposal distribution $q$.
In addition,  it is usually necessary to discard a large number of initial samples (\textit{burn-in}), and the parameter values in the chain will be auto-correlated reducing the effective sample size \cite{brooks2011handbook}.}.

Particle methods, such as sequential Monte Carlo (SMC) and related methods \cite{doucet2001sequential}  remove\footnote{Ironically given their name.} some of the sequential nature of MCMC. Although  most often used in time structured problems,   they can also be used in static problems. As in   importance sampling \cite{robert1999monte}, they aim  to form a weighted set of particles $\{\bx^{(i)}, w^{(i)}\}_{i=1}^N$ so  that any  posterior expectation can be approximated by a weighted sum:
$$\BE [h(\mathbf{X})\mid \by] = \int h(\bx) \pi(\bx \mid \by) \mathrm{d}\bx\approx \sum_{i=1}^N w^{(i)} h(\bx^{(i)}).$$
A  series of target distributions,  $\pi^{(j)}(\bx^* \mid \by)$, is formed that converges to $\pi(\bx^* \mid \by)$ as $j=1, \ldots, J$, where $J$ is typically small. Each stage typically requires $N$ new simulations, which can be run in parallel.
There are various ways to define the targets $\pi^{(j)}$, such as incorporating additional data points, e.g. $\pi^{(j)}(\bx \mid \by) = \pi (\bx^* \mid y_{1:j})$, or requiring greater accuracy by reducing the error term as in tempering and sequential ABC  \cite{sisson2007sequential, doucet2001introduction}.

Unlike MCMC methods, SMC methods are easily parallelizable. However,  particle degeneracy often occurs (where the variance of the weights is large), which in severe cases will mean the sampler needs to be restarted afresh.
Like MCMC methods,  they typically require a huge number of simulator evaluations making them infeasible for many problems, and can easily fail to explore all of the modes in multi-modal posterior distributions.

MCMC and SMC methods are only practicable for  computationally cheap simulators (where `cheap' is relative to your computational resource). In cases where the computational budget is limited, we can often make stronger assumptions somewhere in the analysis and allow for a greater degree of approximation.  For example, methods based on  Kalman inversion \cite{evensen2003ensemble, iglesias2013ensemble} often assume that the posterior can be approximated by a Gaussian distribution $\pi(\bx^* \mid \by) \approx N(\bx^* ; \bmu, \bSigma)$. Gaussian distributions are fully characterized by their mean vector $\bmu$ and covariance matrix $\bSigma$, and these can be estimated with many fewer particles than are needed for non-parametric approximation methods such as MCMC or SMC. This can allow us to approximate the posterior with fewer  simulator evaluations. Of course, if the posterior is not well approximated by a Gaussian distribution, for example if the posterior is multimodal, skewed, or heavy-tailed, then the approximation will be poor.

\subsection{Optimization based approaches}

If we only require a point estimate of the best input, $\bx^*$, then we can use standard optimization algorithms to minimize some cost function such as the log-likelihood.
There are also optimization approaches (as opposed to sampling approaches) for approximating  Bayesian posterior distributions. In variational inference \cite{blei2017variational}, we assume a distributional form for the posterior, e.g. $q_\phi(\bx)$, parameterized by unknown parameters $\phi$. For example, we could  use a mean field approximation with Gaussian marginals
\begin{equation}
q_\phi(\bx^*) =  \prod_{i=1}^d \mathcal{N}(x_i; \mu_i, \sigma_i^2).
\label{eq:meanfield}
\end{equation}
The variational parameters $\phi=(\mu_1,\ldots, \mu_d, \sigma_1^2, \ldots, \sigma_d^2)$  are estimated by minimizing the distance between $q_\phi$ and the posterior $\pi(\bx^*\mid \by)$ with respect to $\phi$.
If this distance is measured with the Kullback-Leibler (KL) divergence, then this results in an optimization problem that requires us to minimize the sum of the expected log-likelihood under $q_\phi$, plus the KL divergence between $q_\phi$ and the prior distribution for $\bx$.
\begin{equation}\min_\phi \left(-\BE_{\bx\sim q_\phi} \log \pi(\by \mid \bx) + KL(q_\phi(\bx) ||\pi(\bx))\right).\label{eq:VI}
\end{equation}
Efficient solution of this problem usually requires derivative information about the simulator
\cite{ranganath2014black}. Eq. \ref{eq:VI}  can be generalized  with a generic score and divergence (i.e., something other than the negative log-likelihood and KL divergence)
$$\min_{q\in\mathcal{Q}} \left(\BE_{\bx\sim q} S(\bx, \bbf, \by) + D(q(\bx) ||\pi(\bx))\right)$$
which gives an alternative way to generalize Bayesian inference  that may help to mitigate the effects of model misspecification \cite{knoblauch2019generalized, matsubara2022robust,masegosa2020learning}. Here, $\mathcal{Q}$ is the space of probability distribution we search in for approximations. If we choose too limited a space, such as Eq. \ref{eq:meanfield}, the resulting approximation may be poor; choose too rich a class, and we may not be able to solve the optimization problem given the computational budget.





In \textit{amortized} Bayesian inference \cite{gershman2014amortized}, we aim to learn  a density estimation network, which once trained,  can be  applied to any new observation $\by$ to quickly give an estimate of $\pi(\bx^* \mid \by)$. For example, rather than fitting a variational posterior $q_\phi(\bx^*)$ that is specific to a given instance of $\by$, we instead learn  a variational approximation of the form $q_\phi(\bx^*\mid \by)$, which holds for {\it any} observed $\by$ \cite{greenberg2019automatic, gonccalves2020training,
papamakarios2016fast}. Typically a neural network will be used for $q_\phi$, resulting in a similar optimization problem as  occurs in  variational auto-encoders and related models \cite{kingma2013auto, rezende2015variational}. 



\subsection{Surrogate model methods}

For many complex simulators, the computational cost, combined with  lack of knowledge about the simulator, mean that we are unable to calibrate with the methods discussed above without using additional simplifying approximations. One of the most widely used approaches is to  use a \textit{surrogate model}, or \textit{emulator}, of the simulator \cite{sacks_design_1989,  sudret2017surrogate, gramacy2020surrogates}. The idea is that if we can find an  approximation, $\hat{f}(\bx)$, of the simulator output at $\bx$ that is fast to evaluate, then we can use $\hat{f}$ instead of $f$ to calibrate the simulator. If we can quantify the accuracy of $\hat{f}(\bx)$ with a probabilistic model we can take this uncertainty about the true value of $\bbf(\bx)$ (\textit{code uncertainty} \cite{o2006bayesian}) into account in the calibration.

There are many different approaches to building surrogate models, the most common being to use a data-driven approximation to exploit the continuity and potential smoothness of the simulator response as a function of the parameter $\bx$.
Suppose we can afford $N$ simulator evaluations, which we collect in the ensemble $\mathcal{D}=\{\bx^{(i)}, f(\bx^{(i)})\}_{i=1}^N$. Using $\mathcal{D}$ we can build a probabilistic model for $f$, $\pi(f \mid \mathcal{D})$, that gives us a predictive distribution for $f(\bx)$ for any $\bx$.
In a Bayesian calibration, we would then try to find the posterior distribution for $\bx$ given data $\by$ and the ensemble $\mathcal{D}$:
\begin{equation}
\pi(\bx^* \mid \by, \D) = \int \pi(\bx^* \mid \by, f) \pi(f \mid \D) {\rm d}f
\end{equation}
Note that before, the posterior was implicitly conditioned on the simulator, i.e., $\pi(\bx^* \mid \by) \equiv \pi(\bx^* \mid, \by, f)$, whereas now it is only conditioned on the ensemble of simulator runs, $\D$, as well as any assumptions made in the surrogate.  We will expect $\pi(\bx^* \mid \by, \D)$ to be more uncertain about the best input than $\pi(\bx^* \mid \by, f)$.

In practice we may sometimes ignore some of the uncertainties, for example, by finding a point estimate of $f$, and then working as if this is correct, i.e., finding $\pi(\bx^* \mid \by, \hat{f})$. Or if we use a Gaussian process surrogate model \cite{rasmussen2006gaussian}  for $f$, then there are often hyper-parameters $\phi$ that parameterize the covariance function,  and these are often estimated and then fixed, so that process uncertainty is integrated out but uncertainty in $\phi$ is ignored, i.e., we use $\pi(\bx \mid \by, \D, \hat{\phi})$. Ignoring some of the uncertainty about $f$ can lead to over-confident posteriors, so care is needed whenever uncertainty is ignored.

We can also decide to build a surrogate of the cost function. For example, one application of surrogates that is particularly well developed is {\it Bayesian optimization} \cite{frazier2018tutorial},
which is useful for finding point estimates (or variational posteriors) when the computational cost of the simulator is high. For example, we might  build a surrogate model, $\hat{S}(\bx)$, for the cost function ${S}(\bx)$, usually using a Gaussian process (GP), as this gives both a point estimate for $S(\bx)$, and an associated uncertainty. 
An acquisition rule is then used to determine  where next to sample the simulator \cite{wilson2018maximizing}, taking into account the uncertainty in $\hat{S}$ in a way that trades off exploration and exploitation. In the exploration phase, the acquisition function tends to suggest values of $\bx$ to fill in gaps in the surrogate model's knowledge, i.e., where $\text{Var}[\hat S(\bx)]$ is large, whereas in the exploitation phase, it suggests  values of $\bx$ that it thinks are near the minimum. Similar ideas have been developed for sampling approaches  for calibration. For example,  in an MCMC sampler, we can introduce an intermediate step between the proposal of a move to $\bx'$, and computing the acceptance ratio, where we first look to see how confident our surrogate is about $f(\bx')$,   and as a result then  decide to either compute the acceptance ratio using $\hat{f}(\bx')$ or to do an additional simulation to find $f(\bx')$ \cite{conrad2016accelerating}.

Although using a surrogate initially sounds like it will introduce an additional error in the calibration, in practice it can result in more accurate inferences \cite{hennig2015probabilistic,hennig2022probabilistic}. If your computational budget allows for $N$ evaluations of $f$, our inference of $\bx^*$ can then either be based on just those $N$ simulations, or we can build a surrogate of $f$ that interpolates and extrapolates from $\D$ to guess at the simulator behaviour at points $\bx$ not in the training set. In other words, we can exploit  the  continuity and smoothness of $f$, resulting in more accurate inferences.
It depends upon the problem specifics whether building a surrogate will be successful or not, but for many problems, there is often an underlying smooth response surface that can be modelled. 








\section{Conclusions}


Although we have presented decisions about the calibration workflow (observational model, framework, computational approach) as if they are taken at the outset and then set in stone, in practice, a degree of pragmatism is often required. After preliminary attempts at calibration, we may need to revise our choices once we are able to  assess what is realistic. It is also important to consider what degree of accuracy is necessary in the calibration.
The accuracy with which you will be able to estimate the best input, will depend upon the accuracy of the underlying mathematical model (which is likely incomplete and misspecified), the numerical solver (which will be approximate), the noise on the data, and the numerical error in your calibration scheme (i.e., the error between the true posterior and your approximation to it). A common mistake is to fixate on minimizing the latter, even when that error will be dominated by other errors in the final approximation.  

\bibliographystyle{plain} 
\bibliography{example_bib} 

\end{document}

%% file: preamble.tex
\newcommand{\BS}[1]{\textcolor{blue}{#1}}
\newcommand{\bbf}{\bm{f}}
\newcommand{\bx}{\bm{x}}
\newcommand{\bd}{\bm{d}}
\newcommand{\bg}{\bm{g}}
\newcommand{\bz}{\bm{z}}
\newcommand{\by}{\bm{y}}
\newcommand{\bw}{\bm{w}}
\newcommand{\bS}{\bm{S}}
\newcommand{\D}{\mathcal{D}}
\newcommand{\bT}{\bm{T}}
\newcommand{\rd}{{\rm{d}}}
\newcommand{\bu}{\bm{u}}
\newcommand{\BE}{E}
\newcommand{\bepsilon}{\bm{\epsilon}}
\newcommand{\bSigma}{\bm{\Sigma}}
\newcommand{\bmu}{\bm{\mu}}

\newcommand{\X}{\mathcal{X}}
\newcommand{\Y}{\mathcal{Y}}
\newcommand{\F}{\mathcal{F}}
\newcommand{\U}{\mathcal{U}}

%% file: main.bbl
\begin{thebibliography}{100}

\bibitem{abadi2016tensorflow}
Mart{\'\i}n Abadi, Paul Barham, Jianmin Chen, Zhifeng Chen, Andy Davis, Jeffrey
  Dean, Matthieu Devin, Sanjay Ghemawat, Geoffrey Irving, Michael Isard, et~al.
\newblock Tensorflow: a system for large-scale machine learning.
\newblock In {\em Osdi}, volume~16, pages 265--283. Savannah, GA, USA, 2016.

\bibitem{arendt2012quantification}
Paul~D. Arendt, Daniel~W. Apley, and Wei Chen.
\newblock {Quantification of Model Uncertainty: Calibration, Model Discrepancy,
  and Identifiability}.
\newblock {\em Journal of Mechanical Design}, 134(10), 09 2012.

\bibitem{arsenault2014comparison}
Richard Arsenault, Annie Poulin, Pascal C{\^o}t{\'e}, and Fran{\c{c}}ois
  Brissette.
\newblock Comparison of stochastic optimization algorithms in hydrological
  model calibration.
\newblock {\em Journal of Hydrologic Engineering}, 19(7):1374--1384, 2014.

\bibitem{bardossy2007calibration}
Andr{\'a}s B{\'a}rdossy.
\newblock Calibration of hydrological model parameters for ungauged catchments.
\newblock {\em Hydrology and Earth System Sciences}, 11(2):703--710, 2007.

\bibitem{beaumont2002approximate}
Mark~A Beaumont, Wenyang Zhang, and David~J Balding.
\newblock Approximate {B}ayesian computation in population genetics.
\newblock {\em Genetics}, 162(4):2025--2035, 2002.

\bibitem{bellprat2012objective}
Omar Bellprat, Sven Kotlarski, Daniel L{\"u}thi, and Christoph Sch{\"a}r.
\newblock Objective calibration of regional climate models.
\newblock {\em Journal of Geophysical Research: Atmospheres}, 117(D23), 2012.

\bibitem{berger1990robust}
James~O Berger.
\newblock Robust {B}ayesian analysis: sensitivity to the prior.
\newblock {\em Journal of statistical planning and inference}, 25(3):303--328,
  1990.

\bibitem{berrar2019cross}
Daniel Berrar.
\newblock Cross-validation.
\newblock In Shoba Ranganathan, Kenta Nakai, and Christian Schonbach, editors,
  {\em Encyclopedia of bioinformatics and computational biology: ABC of
  bioinformatics}, volume~1, pages 542--545. Elsevier, 2018.

\bibitem{beven2001equifinality}
Keith Beven and Jim Freer.
\newblock Equifinality, data assimilation, and uncertainty estimation in
  mechanistic modelling of complex environmental systems using the {GLUE}
  methodology.
\newblock {\em Journal of hydrology}, 249(1-4):11--29, 2001.

\bibitem{bierkens2019zig}
Joris Bierkens, Paul Fearnhead, and Gareth Roberts.
\newblock {The Zig-Zag process and super-efficient sampling for Bayesian
  analysis of big data}.
\newblock {\em The Annals of Statistics}, 47(3):1288 -- 1320, 2019.

\bibitem{bissiri2016general}
Pier~Giovanni Bissiri, Chris~C Holmes, and Stephen~G Walker.
\newblock A general framework for updating belief distributions.
\newblock {\em Journal of the Royal Statistical Society. Series B, Statistical
  Methodology}, 78(5):1103, 2016.

\bibitem{blei2017variational}
David~M Blei, Alp Kucukelbir, and Jon~D McAuliffe.
\newblock Variational inference: A review for statisticians.
\newblock {\em Journal of the American statistical Association},
  112(518):859--877, 2017.

\bibitem{bloem2020probabilistic}
Benjamin Bloem-Reddy and Yee~Whye Teh.
\newblock Probabilistic symmetries and invariant neural networks.
\newblock {\em The Journal of Machine Learning Research}, 21(1):3535--3595,
  2020.

\bibitem{box1976science}
George~EP Box.
\newblock Science and statistics.
\newblock {\em Journal of the American Statistical Association},
  71(356):791--799, 1976.

\bibitem{jax2018github}
James Bradbury, Roy Frostig, Peter Hawkins, Matthew~James Johnson, Chris Leary,
  Dougal Maclaurin, George Necula, Adam Paszke, Jake Vander{P}las, Skye
  Wanderman-{M}ilne, and Qiao Zhang.
\newblock {JAX}: composable transformations of {P}ython+{N}um{P}y programs.
\newblock http://github.com/google/jax, 2018.

\bibitem{brooks2011handbook}
Steve Brooks, Andrew Gelman, Galin Jones, and Xiao-Li Meng.
\newblock {\em Handbook of {M}arkov chain {M}onte {C}arlo}.
\newblock CRC press, 2011.

\bibitem{brynjarsdottir2014learning}
Jenn{\'y} Brynjarsd{\`o}ttir and Anthony O'Hagan.
\newblock Learning about physical parameters: The importance of model
  discrepancy.
\newblock {\em Inverse problems}, 30(11):114007, 2014.

\bibitem{cao2003adjoint}
Yang Cao, Shengtai Li, Linda Petzold, and Radu Serban.
\newblock Adjoint sensitivity analysis for differential-algebraic equations:
  The adjoint {DAE} system and its numerical solution.
\newblock {\em SIAM journal on scientific computing}, 24(3):1076--1089, 2003.

\bibitem{carpenter2017stan}
Bob Carpenter, Andrew Gelman, Matthew~D Hoffman, Daniel Lee, Ben Goodrich,
  Michael Betancourt, Marcus Brubaker, Jiqiang Guo, Peter Li, and Allen
  Riddell.
\newblock Stan: A probabilistic programming language.
\newblock {\em Journal of statistical software}, 76(1), 2017.

\bibitem{carson2019quantifying}
Jake Carson, Michel Crucifix, Simon~P Preston, and Richard~D Wilkinson.
\newblock Quantifying age and model uncertainties in palaeoclimate data and
  dynamical climate models with a joint inferential analysis.
\newblock {\em Proceedings of the Royal Society A}, 475(2224):20180854, 2019.

\bibitem{caruel2014dimensional}
Matthieu Caruel, Radomir Chabiniok, Philippe Moireau, Yves Lecarpentier, and
  Dominique Chapelle.
\newblock Dimensional reductions of a cardiac model for effective validation
  and calibration.
\newblock {\em Biomechanics and modeling in mechanobiology}, 13:897--914, 2014.

\bibitem{chang2019computer}
Kai-Lan Chang and Serge Guillas.
\newblock Computer model calibration with large non-stationary spatial outputs:
  application to the calibration of a climate model.
\newblock {\em Journal of the Royal Statistical Society: Series C (Applied
  Statistics)}, 68(1):51--78, 2019.

\bibitem{chang2016calibrating}
Won Chang, Murali Haran, Patrick Applegate, and David Pollard.
\newblock Calibrating an ice sheet model using high-dimensional binary spatial
  data.
\newblock {\em Journal of the American Statistical Association},
  111(513):57--72, 2016.

\bibitem{chen2018neural}
Ricky~TQ Chen, Yulia Rubanova, Jesse Bettencourt, and David~K Duvenaud.
\newblock Neural ordinary differential equations.
\newblock {\em Advances in neural information processing systems}, 31, 2018.

\bibitem{christen2005markov}
J~Andr{\'e}s Christen and Colin Fox.
\newblock Markov chain {M}onte {C}arlo using an approximation.
\newblock {\em Journal of Computational and Graphical statistics},
  14(4):795--810, 2005.

\bibitem{conrad2016accelerating}
Patrick~R Conrad, Youssef~M Marzouk, Natesh~S Pillai, and Aaron Smith.
\newblock Accelerating asymptotically exact {MCMC} for computationally
  intensive models via local approximations.
\newblock {\em Journal of the American Statistical Association},
  111(516):1591--1607, 2016.

\bibitem{constantine2015active}
Paul~G Constantine.
\newblock {\em Active subspaces: Emerging ideas for dimension reduction in
  parameter studies}.
\newblock SIAM, 2015.

\bibitem{constantine2016accelerating}
Paul~G Constantine, Carson Kent, and Tan Bui-Thanh.
\newblock Accelerating {Markov chain Monte Carlo} with active subspaces.
\newblock {\em SIAM Journal on Scientific Computing}, 38(5):A2779--A2805, 2016.

\bibitem{cotter2013mcmc}
Simon~L Cotter, Gareth~O Roberts, Andrew~M Stuart, and David White.
\newblock {MCMC} methods for functions: modifying old algorithms to make them
  faster.
\newblock {\em Statistical Science}, 28(3):424--446, 2013.

\bibitem{couvreux2021process}
Fleur Couvreux, Fr{\'e}d{\'e}ric Hourdin, Daniel Williamson, Romain Roehrig,
  Victoria Volodina, Najda Villefranque, Catherine Rio, Olivier Audouin, James
  Salter, Eric Bazile, et~al.
\newblock Process-based climate model development harnessing machine learning:
  {I}. a calibration tool for parameterization improvement.
\newblock {\em Journal of Advances in Modeling Earth Systems},
  13(3):e2020MS002217, 2021.

\bibitem{craig1997pressure}
Peter~S Craig, Michael Goldstein, Allan~H Seheult, and James~A Smith.
\newblock Pressure matching for hydrocarbon reservoirs: a case study in the use
  of {B}ayes linear strategies for large computer experiments.
\newblock In {\em Case Studies in {B}ayesian Statistics: Volume III}, pages
  37--93. Springer, 1997.

\bibitem{csillery2010approximate}
Katalin Csill{\'e}ry, Michael~GB Blum, Oscar~E Gaggiotti, and Olivier
  Fran{\c{c}}ois.
\newblock Approximate {B}ayesian computation {(ABC)} in practice.
\newblock {\em Trends in ecology \& evolution}, 25(7):410--418, 2010.

\bibitem{cui2014likelihood}
Tiangang Cui, James Martin, Youssef~M Marzouk, Antti Solonen, and Alessio
  Spantini.
\newblock Likelihood-informed dimension reduction for nonlinear inverse
  problems.
\newblock {\em Inverse Problems}, 30(11):114015, 2014.

\bibitem{davis2007methods}
Philip~J Davis and Philip Rabinowitz.
\newblock {\em Methods of numerical integration}.
\newblock Courier Corporation, 2007.

\bibitem{dellaportas2003introduction}
Petros Dellaportas and Gareth~O Roberts.
\newblock An introduction to {MCMC}.
\newblock In {\em Spatial statistics and computational methods}, pages 1--41.
  Springer, 2003.

\bibitem{dodwell2015hierarchical}
Tim~J Dodwell, Christian Ketelsen, Robert Scheichl, and Aretha~L Teckentrup.
\newblock A hierarchical multilevel {Markov chain Monte Carlo} algorithm with
  applications to uncertainty quantification in subsurface flow.
\newblock {\em SIAM/ASA Journal on Uncertainty Quantification},
  3(1):1075--1108, 2015.

\bibitem{doucet2001introduction}
Arnaud Doucet, Nando De~Freitas, and Neil Gordon.
\newblock An introduction to sequential {Monte Carlo} methods.
\newblock In {\em Sequential Monte Carlo methods in practice}, pages 3--14.
  Springer, 2001.

\bibitem{doucet2001sequential}
Arnaud Doucet, Nando De~Freitas, Neil~James Gordon, et~al.
\newblock {\em Sequential {Monte Carlo} methods in practice}, volume~1.
\newblock Springer, 2001.

\bibitem{edwards2011precalibrating}
Neil~R Edwards, David Cameron, and Jonathan Rougier.
\newblock Precalibrating an intermediate complexity climate model.
\newblock {\em Climate dynamics}, 37:1469--1482, 2011.

\bibitem{estep2004short}
Donald Estep.
\newblock A short course on duality, adjoint operators, {G}reen’s functions,
  and a posteriori error analysis.
\newblock {\em Lecture Notes}, 2004.

\bibitem{evensen2003ensemble}
Geir Evensen.
\newblock The ensemble {K}alman filter: Theoretical formulation and practical
  implementation.
\newblock {\em Ocean dynamics}, 53:343--367, 2003.

\bibitem{frazier2018tutorial}
Peter~I Frazier.
\newblock A tutorial on {B}ayesian optimization.
\newblock {\em arXiv preprint arXiv:1807.02811}, 2018.

\bibitem{frieboes2009prediction}
Hermann~B Frieboes, Mary~E Edgerton, John~P Fruehauf, Felicity~RAJ Rose, Lisa~K
  Worrall, Robert~A Gatenby, Mauro Ferrari, and Vittorio Cristini.
\newblock Prediction of drug response in breast cancer using integrative
  experimental/computational modeling.
\newblock {\em Cancer research}, 69(10):4484--4492, 2009.

\bibitem{gahungu2022adjoint}
Paterne Gahungu, Christopher~W Lanyon, Mauricio~A \'{A}lvarez, Engineer
  Bainomugisha, Michael~T Smith, and Richard~D Wilkinson.
\newblock Adjoint-aided inference of {G}aussian process driven differential
  equations.
\newblock {\em Advances in Neural Information Processing Systems},
  35:17233--17247, 2022.

\bibitem{gelfand1990sampling}
Alan~E Gelfand and Adrian~FM Smith.
\newblock Sampling-based approaches to calculating marginal densities.
\newblock {\em Journal of the American statistical association},
  85(410):398--409, 1990.

\bibitem{gelman2017prior}
Andrew Gelman, Daniel Simpson, and Michael Betancourt.
\newblock The prior can often only be understood in the context of the
  likelihood.
\newblock {\em Entropy}, 19(10):555, 2017.

\bibitem{gershman2014amortized}
Samuel Gershman and Noah Goodman.
\newblock Amortized inference in probabilistic reasoning.
\newblock In {\em Proceedings of the annual meeting of the cognitive science
  society}, volume~36, 2014.

\bibitem{ginsbourger2016degeneracy}
David Ginsbourger, Olivier Roustant, and Nicolas Durrande.
\newblock On degeneracy and invariances of random fields paths with
  applications in {G}aussian process modelling.
\newblock {\em Journal of statistical planning and inference}, 170:117--128,
  2016.

\bibitem{girolami2011riemann}
Mark Girolami and Ben Calderhead.
\newblock Riemann manifold {L}angevin and {H}amiltonian {M}onte {C}arlo
  methods.
\newblock {\em Journal of the Royal Statistical Society: Series B (Statistical
  Methodology)}, 73(2):123--214, 2011.

\bibitem{goh2013prediction}
Joslin Goh, Derek Bingham, James~Paul Holloway, Michael~J Grosskopf, Carolyn~C
  Kuranz, and Erica Rutter.
\newblock Prediction and computer model calibration using outputs from
  multifidelity simulators.
\newblock {\em Technometrics}, 55(4):501--512, 2013.

\bibitem{gonccalves2020training}
Pedro~J Gon{\c{c}}alves, Jan-Matthis Lueckmann, Michael Deistler, Marcel
  Nonnenmacher, Kaan {\"O}cal, Giacomo Bassetto, Chaitanya Chintaluri,
  William~F Podlaski, Sara~A Haddad, Tim~P Vogels, et~al.
\newblock Training deep neural density estimators to identify mechanistic
  models of neural dynamics.
\newblock {\em Elife}, 9:e56261, 2020.

\bibitem{goodfellow2016deep}
Ian Goodfellow, Yoshua Bengio, and Aaron Courville.
\newblock {\em Deep learning}.
\newblock MIT press, 2016.

\bibitem{gosling2018shelf}
John~Paul Gosling.
\newblock {SHELF}: the {S}heffield elicitation framework.
\newblock In {\em Elicitation: The science and art of structuring judgement},
  pages 61--93. Springer, 2018.

\bibitem{gramacy2020surrogates}
Robert~B Gramacy.
\newblock {\em Surrogates: Gaussian process modeling, design, and optimization
  for the applied sciences}.
\newblock CRC press, 2020.

\bibitem{greenberg2019automatic}
David Greenberg, Marcel Nonnenmacher, and Jakob Macke.
\newblock Automatic posterior transformation for likelihood-free inference.
\newblock In {\em International Conference on Machine Learning}, pages
  2404--2414. PMLR, 2019.

\bibitem{grunwald2021pac}
Peter Grunwald, Thomas Steinke, and Lydia Zakynthinou.
\newblock {PAC-Bayes, MAC-Bayes} and conditional mutual information: Fast rate
  bounds that handle general {VC} classes.
\newblock In {\em Conference on Learning Theory}, pages 2217--2247. PMLR, 2021.

\bibitem{grunwald2017inconsistency}
Peter Gr{\"u}nwald and Thijs Van~Ommen.
\newblock Inconsistency of {B}ayesian inference for misspecified linear models,
  and a proposal for repairing it.
\newblock 2017.

\bibitem{gupta1998toward}
Hoshin~Vijai Gupta, Soroosh Sorooshian, and Patrice~Ogou Yapo.
\newblock Toward improved calibration of hydrologic models: Multiple and
  noncommensurable measures of information.
\newblock {\em Water Resources Research}, 34(4):751--763, 1998.

\bibitem{guzman2010genetic}
R~Guzm{\'a}n-Cruz, R~Casta{\~n}eda-Miranda, JJ~Garc{\'\i}a-Escalante,
  A~Lara-Herrera, I~Serroukh, and LO~Solis-S{\'a}nchez.
\newblock Genetic algorithms for calibration of a greenhouse climate model.
\newblock {\em Revista Chapingo. Serie horticultura}, 16(1):23--30, 2010.

\bibitem{hastie2009elements}
Trevor Hastie, Robert Tibshirani, Jerome~H Friedman, and Jerome~H Friedman.
\newblock {\em The elements of statistical learning: data mining, inference,
  and prediction}, volume~2.
\newblock Springer, 2009.

\bibitem{hennig2015probabilistic}
Philipp Hennig, Michael~A Osborne, and Mark Girolami.
\newblock Probabilistic numerics and uncertainty in computations.
\newblock {\em Proceedings of the Royal Society A: Mathematical, Physical and
  Engineering Sciences}, 471(2179):20150142, 2015.

\bibitem{hennig2022probabilistic}
Philipp Hennig, Michael~A Osborne, and Hans~P Kersting.
\newblock {\em Probabilistic Numerics: Computation as Machine Learning}.
\newblock Cambridge University Press, 2022.

\bibitem{hoffman2014no}
Matthew~D Hoffman, Andrew Gelman, et~al.
\newblock The {No-U-Turn} sampler: adaptively setting path lengths in
  {Hamiltonian Monte Carlo}.
\newblock {\em J. Mach. Learn. Res.}, 15(1):1593--1623, 2014.

\bibitem{holden2018abc}
Philip~B Holden, Neil~R Edwards, James Hensman, and Richard~D Wilkinson.
\newblock {ABC} for climate: dealing with expensive simulators.
\newblock In {\em Handbook of approximate {B}ayesian computation}, pages
  569--595. Chapman and Hall/CRC, 2018.

\bibitem{hourdin2017art}
Fr{\'e}d{\'e}ric Hourdin, Thorsten Mauritsen, Andrew Gettelman, Jean-Christophe
  Golaz, Venkatramani Balaji, Qingyun Duan, Doris Folini, Duoying Ji, Daniel
  Klocke, Yun Qian, Florian Rauser, Catherine Rio, Lorenzo Tomassini, Masahiro
  Watanabe, and Daniel Williamson.
\newblock The art and science of climate model tuning.
\newblock {\em Bulletin of the American Meteorological Society},
  98(3):589--602, 2017.

\bibitem{hourdin2021process}
Fr{\'e}d{\'e}ric Hourdin, Daniel Williamson, Catherine Rio, Fleur Couvreux,
  Romain Roehrig, Najda Villefranque, Ionela Musat, Laurent Fairhead, F~Binta
  Diallo, and Victoria Volodina.
\newblock Process-based climate model development harnessing machine learning:
  Ii. model calibration from single column to global.
\newblock {\em Journal of Advances in Modeling Earth Systems},
  13(6):e2020MS002225, 2021.

\bibitem{huang2006global}
Deng Huang, Theodore~T Allen, William~I Notz, Ning Zeng, et~al.
\newblock Global optimization of stochastic black-box systems via sequential
  kriging meta-models.
\newblock {\em Journal of global optimization}, 34(3):441--466, 2006.

\bibitem{huang2019tuning}
Shih-Ting Huang, Yannick D{\"u}ren, Kristoffer~H Hellton, and Johannes Lederer.
\newblock Tuning parameter calibration for prediction in personalized medicine.
\newblock {\em arXiv preprint arXiv:1909.10635}, 2019.

\bibitem{iglesias2013ensemble}
Marco~A Iglesias, Kody~JH Law, and Andrew~M Stuart.
\newblock Ensemble kalman methods for inverse problems.
\newblock {\em Inverse Problems}, 29(4):045001, 2013.

\bibitem{jewson2018principles}
Jack Jewson, Jim~Q Smith, and Chris Holmes.
\newblock Principles of {B}ayesian inference using general divergence criteria.
\newblock {\em Entropy}, 20(6):442, 2018.

\bibitem{joseph2009statistical}
V~Roshan Joseph and Shreyes~N Melkote.
\newblock Statistical adjustments to engineering models.
\newblock {\em Journal of Quality Technology}, 41(4):362--375, 2009.

\bibitem{joyce2008approximately}
Paul Joyce and Paul Marjoram.
\newblock Approximately sufficient statistics and {B}ayesian computation.
\newblock {\em Statistical applications in genetics and molecular biology},
  7(1), 2008.

\bibitem{kennedy_bayesian_2001}
M.~C. Kennedy and A.~O'Hagan.
\newblock Bayesian calibration of computer models.
\newblock {\em Journal of the Royal Statistical Society: Series B (Statistical
  Methodology)}, 63(3):425--464, January 2001.

\bibitem{kingma2013auto}
Diederik~P. Kingma and Max Welling.
\newblock {Auto-Encoding Variational Bayes}.
\newblock In {\em 2nd International Conference on Learning Representations,
  {ICLR} 2014, Banff, AB, Canada, April 14-16, 2014, Conference Track
  Proceedings}, 2014.

\bibitem{knoblauch2019generalized}
Jeremias Knoblauch, Jack Jewson, and Theodoros Damoulas.
\newblock Generalized variational inference: Three arguments for deriving new
  posteriors.
\newblock {\em arXiv preprint arXiv:1904.02063}, 2019.

\bibitem{large2001equatorial}
William~G Large, Gokhan Danabasoglu, James~C McWilliams, Peter~R Gent, and
  Frank~O Bryan.
\newblock Equatorial circulation of a global ocean climate model with
  anisotropic horizontal viscosity.
\newblock {\em Journal of Physical Oceanography}, 31(2):518--536, 2001.

\bibitem{lunn2012bugs}
David Lunn, Chris Jackson, Nicky Best, Andrew Thomas, and David Spiegelhalter.
\newblock {\em The BUGS book: A practical introduction to {B}ayesian analysis}.
\newblock CRC press, 2012.

\bibitem{march2012constrained}
Andrew March and Karen Willcox.
\newblock Constrained multifidelity optimization using model calibration.
\newblock {\em Structural and Multidisciplinary Optimization}, 46:93--109,
  2012.

\bibitem{marchesseau2013fast}
St{\'e}phanie Marchesseau, Herv{\'e} Delingette, Maxime Sermesant, and Nicholas
  Ayache.
\newblock Fast parameter calibration of a cardiac electromechanical model from
  medical images based on the unscented transform.
\newblock {\em Biomechanics and modeling in mechanobiology}, 12(4):815--831,
  2013.

\bibitem{marin2012approximate}
Jean-Michel Marin, Pierre Pudlo, Christian~P Robert, and Robin~J Ryder.
\newblock Approximate {B}ayesian computational methods.
\newblock {\em Statistics and computing}, 22(6):1167--1180, 2012.

\bibitem{masegosa2020learning}
Andres Masegosa.
\newblock Learning under model misspecification: Applications to variational
  and ensemble methods.
\newblock {\em Advances in Neural Information Processing Systems},
  33:5479--5491, 2020.

\bibitem{matsubara2022robust}
Takuo Matsubara, Jeremias Knoblauch, Fran{\c{c}}ois-Xavier Briol, and Chris~J
  Oates.
\newblock Robust generalised {B}ayesian inference for intractable likelihoods.
\newblock {\em Journal of the Royal Statistical Society Series B: Statistical
  Methodology}, 84(3):997--1022, 2022.

\bibitem{mcmaster1997growing}
Gregory~S McMaster and WW~Wilhelm.
\newblock Growing degree-days: one equation, two interpretations.
\newblock {\em Agricultural and forest meteorology}, 87(4):291--300, 1997.

\bibitem{neal2003slice}
Radford~M Neal.
\newblock Slice sampling.
\newblock {\em The annals of statistics}, 31(3):705--767, 2003.

\bibitem{neal2012bayesian}
Radford~M Neal.
\newblock {\em Bayesian learning for neural networks}, volume 118.
\newblock Springer Science \& Business Media, 2012.

\bibitem{o2006uncertain}
Anthony O'Hagan, Caitlin~E Buck, Alireza Daneshkhah, J~Richard Eiser, Paul~H
  Garthwaite, David~J Jenkinson, Jeremy~E Oakley, and Tim Rakow.
\newblock Uncertain judgements: eliciting experts' probabilities.
\newblock 2006.

\bibitem{o2006bayesian}
Anthony O’Hagan.
\newblock Bayesian analysis of computer code outputs: A tutorial.
\newblock {\em Reliability Engineering \& System Safety}, 91(10-11):1290--1300,
  2006.

\bibitem{papamakarios2016fast}
George Papamakarios and Iain Murray.
\newblock Fast $\varepsilon$-free inference of simulation models with
  {B}ayesian conditional density estimation.
\newblock {\em Advances in neural information processing systems}, 29, 2016.

\bibitem{paszke2017automatic}
Adam Paszke, Sam Gross, Soumith Chintala, Gregory Chanan, Edward Yang, Zachary
  DeVito, Zeming Lin, Alban Desmaison, Luca Antiga, and Adam Lerer.
\newblock Automatic differentiation in pytorch.
\newblock In {\em NIPS 2017 Autodiff Workshop: The Future of Gradient-based
  Machine Learning Software and Techniques}, 2017.

\bibitem{plumlee2017bayesian}
Matthew Plumlee.
\newblock Bayesian calibration of inexact computer models.
\newblock {\em Journal of the American Statistical Association},
  112(519):1274--1285, 2017.

\bibitem{plumlee2016calibrating}
Matthew Plumlee, V~Roshan Joseph, and Hui Yang.
\newblock Calibrating functional parameters in the ion channel models of
  cardiac cells.
\newblock {\em Journal of the American Statistical Association},
  111(514):500--509, 2016.

\bibitem{prangle2015summary}
Dennis Prangle.
\newblock Summary statistics in approximate {B}ayesian computation.
\newblock {\em arXiv preprint arXiv:1512.05633}, 2015.

\bibitem{pukelsheim1994three}
Friedrich Pukelsheim.
\newblock The three sigma rule.
\newblock {\em The American Statistician}, 48(2):88--91, 1994.

\bibitem{raissi2018hidden}
Maziar Raissi and George~Em Karniadakis.
\newblock Hidden physics models: Machine learning of nonlinear partial
  differential equations.
\newblock {\em Journal of Computational Physics}, 357:125--141, 2018.

\bibitem{raissi2019physics}
Maziar Raissi, Paris Perdikaris, and George~E Karniadakis.
\newblock Physics-informed neural networks: A deep learning framework for
  solving forward and inverse problems involving nonlinear partial differential
  equations.
\newblock {\em Journal of Computational physics}, 378:686--707, 2019.

\bibitem{ramamoorthi2015posterior}
RV~Ramamoorthi, Karthik Sriram, and Ryan Martin.
\newblock On posterior concentration in misspecified models.
\newblock {\em Bayesian Analysis}, 10(4):759--789, 2015.

\bibitem{ranganath2014black}
Rajesh Ranganath, Sean Gerrish, and David Blei.
\newblock Black box variational inference.
\newblock In {\em Artificial intelligence and statistics}, pages 814--822.
  PMLR, 2014.

\bibitem{rasmussen2003gaussian}
Carl~Edward Rasmussen.
\newblock Gaussian processes to speed up hybrid {Monte Carlo} for expensive
  {B}ayesian integrals.
\newblock In {\em Seventh Valencia international meeting, dedicated to Dennis
  V. Lindley}, pages 651--659. Oxford University Press, 2003.

\bibitem{rasmussen2006gaussian}
Carl~Edward Rasmussen, Christopher~KI Williams, et~al.
\newblock {\em Gaussian processes for machine learning}, volume~1.
\newblock Springer, 2006.

\bibitem{rezende2015variational}
Danilo Rezende and Shakir Mohamed.
\newblock Variational inference with normalizing flows.
\newblock In {\em International conference on machine learning}, pages
  1530--1538. PMLR, 2015.

\bibitem{rivero2013time}
Cristian~Rodriguez Rivero, Julian Pucheta, Martin Herrera, Victor Sauchelli,
  and Sergio Laboret.
\newblock Time series forecasting using {B}ayesian method: Application to
  cumulative rainfall.
\newblock {\em IEEE Latin America Transactions}, 11(1):359--364, 2013.

\bibitem{robert1999monte}
Christian~P Robert, George Casella, and George Casella.
\newblock {\em Monte Carlo statistical methods}, volume~2.
\newblock Springer, 1999.

\bibitem{roberts1998optimal}
Gareth~O Roberts and Jeffrey~S Rosenthal.
\newblock Optimal scaling of discrete approximations to {L}angevin diffusions.
\newblock {\em Journal of the Royal Statistical Society: Series B (Statistical
  Methodology)}, 60(1):255--268, 1998.

\bibitem{rodero2023calibration}
Cristobal Rodero, Stefano Longobardi, Christoph Augustin, Marina Strocchi,
  Gernot Plank, Pablo Lamata, and Steven~A Niederer.
\newblock Calibration of cohorts of virtual patient heart models using
  {B}ayesian history matching.
\newblock {\em Annals of Biomedical Engineering}, 51(1):241--252, 2023.

\bibitem{sacks_design_1989}
J.~Sacks, W.~J. Welch, T.~J. Mitchell, and H.~P. Wynn.
\newblock Design and {Analysis} of {Computer} {Experiments}.
\newblock {\em Statistical Science}, 4(4):409--423, 1989.

\bibitem{schaeffer2017learning}
Hayden Schaeffer.
\newblock Learning partial differential equations via data discovery and sparse
  optimization.
\newblock {\em Proceedings of the Royal Society A: Mathematical, Physical and
  Engineering Sciences}, 473(2197):20160446, 2017.

\bibitem{shafer2008tutorial}
Glenn Shafer and Vladimir Vovk.
\newblock A tutorial on conformal prediction.
\newblock {\em Journal of Machine Learning Research}, 9(3), 2008.

\bibitem{sisson2018handbook}
Scott~A Sisson, Yanan Fan, and Mark Beaumont.
\newblock {\em Handbook of approximate {B}ayesian computation}.
\newblock CRC Press, 2018.

\bibitem{sisson2007sequential}
Scott~A Sisson, Yanan Fan, and Mark~M Tanaka.
\newblock Sequential {M}onte {C}arlo without likelihoods.
\newblock {\em Proceedings of the National Academy of Sciences},
  104(6):1760--1765, 2007.

\bibitem{sudret2017surrogate}
Bruno Sudret, Stefano Marelli, and Joe Wiart.
\newblock Surrogate models for uncertainty quantification: An overview.
\newblock In {\em 2017 11th European conference on antennas and propagation
  (EUCAP)}, pages 793--797. IEEE, 2017.

\bibitem{sunnaaker2013approximate}
Mikael Sunn{\aa}ker, Alberto~Giovanni Busetto, Elina Numminen, Jukka Corander,
  Matthieu Foll, and Christophe Dessimoz.
\newblock Approximate {B}ayesian computation.
\newblock {\em PLoS computational biology}, 9(1):e1002803, 2013.

\bibitem{tavare1997inferring}
Simon Tavar{\'e}, David~J Balding, Robert~C Griffiths, and Peter Donnelly.
\newblock Inferring coalescence times from {DNA} sequence data.
\newblock {\em Genetics}, 145(2):505--518, 1997.

\bibitem{tuo2015efficient}
Rui Tuo and CF~Jeff Wu.
\newblock Efficient calibration for imperfect computer models.
\newblock {\em Ann. Statist.}, 43(6):2331--2352, 2015.

\bibitem{uteva2017interpolation}
Elena Uteva, Richard~S Graham, Richard~D Wilkinson, and Richard~J Wheatley.
\newblock Interpolation of intermolecular potentials using {G}aussian
  processes.
\newblock {\em The Journal of Chemical Physics}, 147(16):161706, 2017.

\bibitem{van2000asymptotic}
Aad~W Van~der Vaart.
\newblock {\em Asymptotic statistics}, volume~3.
\newblock Cambridge university press, 2000.

\bibitem{vanni2011calibrating}
Tazio Vanni, Jonathan Karnon, Jason Madan, Richard~G White, W~John Edmunds,
  Anna~M Foss, and Rosa Legood.
\newblock Calibrating models in economic evaluation: a seven-step approach.
\newblock {\em Pharmacoeconomics}, 29:35--49, 2011.

\bibitem{vecchia1987simultaneous}
Aldo~V Vecchia and Richard~L Cooley.
\newblock Simultaneous confidence and prediction intervals for nonlinear
  regression models with application to a groundwater flow model.
\newblock {\em Water Resources Research}, 23(7):1237--1250, 1987.

\bibitem{vernon2014galaxy}
Ian Vernon, Michael Goldstein, and Richard Bower.
\newblock Galaxy formation: {B}ayesian history matching for the observable
  universe.
\newblock {\em Statistical science}, pages 81--90, 2014.

\bibitem{whittaker2020calibration}
Dominic~G Whittaker, Michael Clerx, Chon~Lok Lei, David~J Christini, and Gary~R
  Mirams.
\newblock Calibration of ionic and cellular cardiac electrophysiology models.
\newblock {\em Wiley Interdisciplinary Reviews: Systems Biology and Medicine},
  12(4):e1482, 2020.

\bibitem{wilkinson2014accelerating}
Richard Wilkinson.
\newblock Accelerating {ABC} methods using {G}aussian processes.
\newblock In {\em Artificial Intelligence and Statistics}, pages 1015--1023.
  PMLR, 2014.

\bibitem{wilkinson2011quantifying}
Richard~D Wilkinson, Michail Vrettas, Dan Cornford, and Jeremy~E Oakley.
\newblock Quantifying simulator discrepancy in discrete-time dynamical
  simulators.
\newblock {\em Journal of agricultural, biological, and environmental
  statistics}, 16:554--570, 2011.

\bibitem{wilkinson2013approximate}
Richard~David Wilkinson.
\newblock Approximate {B}ayesian computation ({ABC}) gives exact results under
  the assumption of model error.
\newblock {\em Statistical applications in genetics and molecular biology},
  12(2):129--141, 2013.

\bibitem{williamson2013history}
Daniel Williamson, Michael Goldstein, Lesley Allison, Adam Blaker, Peter
  Challenor, Laura Jackson, and Kuniko Yamazaki.
\newblock History matching for exploring and reducing climate model parameter
  space using observations and a large perturbed physics ensemble.
\newblock {\em Climate dynamics}, 41:1703--1729, 2013.

\bibitem{wilson2018maximizing}
James Wilson, Frank Hutter, and Marc Deisenroth.
\newblock Maximizing acquisition functions for {B}ayesian optimization.
\newblock {\em Advances in neural information processing systems}, 31, 2018.

\bibitem{wong2017frequentist}
Raymond~KW Wong, Curtis~B Storlie, and Thomas~CM Lee.
\newblock A frequentist approach to computer model calibration.
\newblock {\em Journal of the Royal Statistical Society. Series B (Statistical
  Methodology)}, pages 635--648, 2017.

\bibitem{yang2018bayesian}
Ziheng Yang and Tianqi Zhu.
\newblock Bayesian selection of misspecified models is overconfident and may
  cause spurious posterior probabilities for phylogenetic trees.
\newblock {\em Proceedings of the National Academy of Sciences},
  115(8):1854--1859, 2018.

\bibitem{zahm2022certified}
Olivier Zahm, Tiangang Cui, Kody Law, Alessio Spantini, and Youssef Marzouk.
\newblock Certified dimension reduction in nonlinear {B}ayesian inverse
  problems.
\newblock {\em Mathematics of Computation}, 91(336):1789--1835, 2022.

\end{thebibliography}
